\documentstyle[12pt]{article}
\begin{document}
\baselineskip=0.6cm
\title{Effect of Landauer's blowtorch on the equilibration rate in a 
bistable potential} 
\author{
Mulugeta Bekele$^{1}$, S. Rajesh$^{2}$, G. Ananthakrishna$^{2}$
and N. Kumar$^{3,4}$\\ $^1$Department of Physics, Addis Ababa
University,\\ Addis Ababa, Ethiopia\\ $^2$Materials Research
Centre, Indian Institute of Science,\\ Bangalore 560 012,
India\\ $^3$Raman Research Institute, C. V. Raman Avenue,\\
Bangalore 560 080, India\\ $^4$Department of Physics, Indian
Institute of Science,\\ Bangalore 560 012, India }
\date{}
\maketitle
\begin{abstract}
\normalsize{ {\em
Kinetic aspect of Landauer's blowtorch effect is investigated
for a model double-well potential with localized heating. Using
the supersymmetric approach, we derive an approximate analytical
expression for the equilibration rate as function of the
strength, width and the position of the hot zone, and the
barrier height. We find that the presence of the hot zone
enhances the equilibration rate, which is found to be an
increasing function of the strength and width of the hot zone.
Our calculations also reveal an intriguing result, namely, that
placing the hot zone away from the top of the potential barrier
enhances the rate more than when it is placed close to it.  A
physically plausible explanation for this is attempted.  The
above analytical results are borne out by detailed numerical
solution of the associated Smoluchowski equation for the
inhomogeneous medium.  }}
\end{abstract}

\section {Introduction}

In a now influential paper \cite{land75} on the relative
stability, {\it i.e.}, relative occupation, of the competing local
energy minima for a system far from equilibrium, Landauer
pointed out the globally determining role played by the
non-equilibrium kinetics of the unstable intermediate states
even as these are very rarely populated.  More specifically, for
the case of two local energy minima, {\it i.e.}, a bistable potential,
he showed that the application of a localized heating at a point
on the reaction coordinate lying between the lower energy
minimum and the potential barrier maximum can raise the relative
population of the higher lying energy minimum over that given by
the thermal Boltzmann factor $exp ( - \Delta\, E / k_B T )$.
This is the so called `blowtorch' effect \cite{land75}
associated with a nonuniform thermal bath.  It generalizes the
problem of escape of a Brownian particle over a potential
barrier under the influence of equilibrium thermal fluctuations,
studied originally by Kramers [2-4], to the case of nonuniform
temperature along the reaction coordinate. Later, in a related
context, it has been shown that a state-dependent diffusion
coefficient can produce maxima in the probability distribution
at points which are not the potential minima [5-7].\\

\noindent
A decade later, van Kampen \cite{vkamp} derived an equation
appropriate for the description of diffusion in an inhomogeneous
medium where he developed a stochastic treatment for the case of
non-uniform temperature. Apart from justifying Landauer's
conjecture, he also showed that there could be a net current
when the particles are allowed to diffuse back through an
alternate route bypassing the hot zone.  The problem has also
been treated by B$\ddot{u}$ttiker around the same time
\cite{butt}. He showed that a net current is possible, even in the absence of an 
externally applied field, provided both the potential and the
state-dependent diffusion constant are periodic with a relative
phase difference.  Later Landauer
\cite{land1} considered this aspect of the problem again in the light of 
van Kampen's work.  Sinha and Moss \cite{sinh} have verified
Landauer's conjecture by computer simulation. This has also been
applied to thermal activation in a superconducting ring with a
weak link where transitions produce temperature changes
\cite{bol}.  Indeed, a whole new field
of research has emerged centering on the idea of the possibility
of directed motion out of noisy states under diverse athermal
driving conditions, generically subsumed under `thermal
ratchets' [13-18], and traceable to the original `blowtorch
theorem' of Landauer in the sense that the latter may be viewed
as injection of noise \cite{mill}.\\

\noindent
Most of the investigations based on the blowtorch effect study
the influence of space-dependent temperature on the steady-state
relative occupations of the energy minima. To the best of
authors' knowledge, there has been no analytical attempt to
study the kinetic aspect of the system, specifically, the
calculation of the longest of relaxation time in a bistable
potential in the presence blowtorch (hot zone).  In this work,
we address this problem and calculate the equilibration rate for
a simple bistable potential with the localized heating
positioned somewhere in the (unstable) intermediate region. The
associated Smoluchowski equation describing the physical
situation will be dealt with using the supersymmetric (SUSY)
approach [20-22]. In doing so, we have adapted the supersymmetric 
potential approach normally applicable to the
original Kramers barrier-crossing problem, to the case when the
temperature is nonuniform along the reaction coordinate.\\

\noindent
The rest of the paper is organized as follows. Section 2 contains 
a brief introduction to the SUSY method for extracting the 
longest relaxation rate ({\it i.e.}, equilibration rate) from the
Smoluchowski equation for the case where temperature is
space-dependent. In section 3, applying the SUSY method to a
model double-well potential with a hot zone, we obtain a matrix
equation determining the equilibration rate. In section 4, we
present an approximate analytical expression obtained from the
matrix equation. The latter is also solved numerically for the
sake of comparison. As an independent check for these results,
we have solved the Smoluchowski equation numerically and
obtained the equilibration rate. The results we obtained from
the above three ways for the equilibration rate as a function of
different parameters characterizing the hot zone are, then,
discussed. We devote the last section to summary and
conclusions.

\section{The methodology}

The Smoluchowski equation describing the kinetics of a Brownian
particle in an inhomogeneous medium is given by
\cite{vkamp}
\begin{equation}
\frac{\partial P(x,t)}{\partial t} =\frac{\partial}{\partial x}
\left[ \mu(x)\left[U^{\prime}(x) P(x,t) + 
\frac{\partial}{\partial x}\left(T(x) P(x,t)\right)\right] \right]
\label{gense}
\end{equation}
\noindent
where the mobility, $\mu$, and temperature, $T$, are in general
space-dependent. (We take Boltzmann's constant, $k_B=1$.) 
Here, $P(x,t)$ is the probability density of
finding the particle at position $x$ at time $t$ and $U(x)$
describes the potential profile and the prime on $U$ denotes the
derivative with respect to $x$.\\

\noindent
Some remarks on the use of the Smoluchowski equation,
Eq.($\ref{gense}$), for the case of inhomogeneous medium are in
order at this point as the latter has been a subject of some
debate that still continues \cite{vkamp,land1,mill,land83,vkamp87}.
Landauer has argued \cite{land1,mill} for a generalization of the 
Smoluchowski equation where the second term on the RHS of 
Eq.($\ref{gense}$) is to be replaced by 
$P(x,t)\left( \frac{\partial T(x)}{\partial x}
\right) + \alpha T(x) \left( \frac{\partial P(x,t)}{\partial x} \right)$. 
The parameter $\alpha$ was shown to depend on the physical
conditions to be imposed across the temperature discontinuity
for no net current.  The parameter $\alpha = 0.5$ corresponds to
a particle interacting with the thermal bath for which the
particle velocity is taken to be proportional to $\sqrt{T}$ or
$P(x) \propto \frac{1}{\sqrt{T(x)}}$. On the other hand, $\alpha
= 1$ corresponds to the case when particles equilibrate via
collision (pressure equilibration), {\it i.e.}, $P(x) \propto
\frac{1}{T(x)}$.
However, a direct derivation based on phase space Smoluchowski
equation by van Kampen gave Eq.($\ref{gense}$) corresponding to
$\alpha = 1$.  This is also supported by the work of Jayannavar
and Mahato \cite{jaya} based on a microscopic treatment of the
thermal bath as a set of harmonic oscillators.  In our work, we
will continue to use Eq.($\ref{gense}$) as providing a
physically valid description of the problem under
consideration.\\

\noindent
We consider the case where temperature is space-dependent but
the medium is homogeneous so that the mobility, $\mu$, is taken
to be constant. Similar assumption of constant $\mu$ has been
taken by Sinha and Moss in their work $\cite{sinh}$. Then, the
corresponding Smoluchowski equation is
\begin{equation}
\frac{\partial P(x,t)}{\partial t} = \mu\frac{\partial}{\partial x}
\left[U^{\prime}(x)P(x,t) + 
\frac{\partial}{\partial x}\left(T(x) P(x,t)\right)\right].
\label{SE}
\end{equation}
\noindent
The stationary solution of Eq.($\ref{SE}$) is given by
\begin{equation}
P_{ss} (x)=\frac{C}{T(x)}
exp\left[-\int^x_{-\infty}\frac{U'(\tilde{x})}{T(\tilde{x})}d\tilde{x}\right],
\label{Pss}
\end{equation}
\noindent
where $C$ is the normalization constant. In this work, we choose 
a simple temperature profile whose value is {\it constant} both 
outside and inside the hot zone with $T = T_0$ and $T = T_0 + T_b$ 
respectively, where $T_b$ is the excess temperature above the 
constant background value $T_0$. (See Eq.($\ref{tmp}$).) Using 
the following transformation
\noindent
\begin{equation}
P(x,t) = \frac{\phi(x)}{T(x)} exp \left[ -\int^x_{-\infty}
\frac{U^{\prime} \, (\tilde{x})}
{2T(\tilde{x})} \, d\tilde{x} \right] e^{- \lambda t}
\label{ansatz}
\end{equation}
\noindent
we convert Eq.($\ref{SE}$) to a Euclidean
Schr$\ddot{o}$dinger equation
\begin{equation}
H_+ \phi_+ = A^+ A \phi_+ = E_+ \phi_+.
\label{schroeq}
\end{equation}
\noindent
Here, $E_+=\frac{\lambda}{\mu T_0}$ and the operators $A$ and
$A^+$ are given by
\begin{equation}
A =\frac{f\partial}{\partial x} +
\frac{U^{\prime}(x)}{2fT_0}\,\,, \hspace{1cm}
A^+ =-\frac{f\partial}{\partial x} + \frac{U^{\prime}(x)}
{2fT_0}.
\label{opera}
\end{equation}

\noindent
In Eq.($\ref{opera}$), the parameter $f$ reflects the excess
temperature defined by $f=\sqrt{1+s}$, with
$s=\frac{T_b}{T_0}$. Thus, for the interval of $x$ outside the hot 
zone, the corresponding operators have $f=1$.  The Hamiltonian 
$H_+$ corresponds to the motion of a particle in the potential
\begin{equation}
V_+(x)=\left(\frac{U^{\prime}(x)}{2T(x)}\right)^2 -
\frac{U^{\prime\prime}(x)}{2T(x)}.
\end{equation}
\noindent
For a high barrier, the longest relaxation rate is determined by
the smallest nonzero eigenvalue $\lambda_1$ associated with the
first excited state of Eq.($\ref{schroeq}$).  On the other hand,
this eigenstate is degenerate with the ground state $\phi^0_-$
of the `supersymmetric partner potential' $V_-(x)$ given by
\begin{equation}
V_-(x)=\left(\frac{U^{\prime}(x)}{2T(x)}\right)^2 +
\frac{U^{\prime\prime}(x)}{2T(x)},
\end{equation} 
\noindent
such that
\begin{equation}
H_-\phi_-^0=AA^+\phi_-^0=E_-\phi_-^0,
\end{equation}
\noindent
where $E_-=\frac{\lambda_1}{\mu T_0}$. Thus, the problem of
finding the equilibration rate amounts to finding the ground
state eigenvalue of this `partner' potential. It is worth
pointing out here that for the temperature profile we use, we
{\it exclude} the two points where $T(x)$ is discontinuous.
However, following the line of attack used for obtaining the
solution of the Schr$\ddot{o}$ndinger equation in the presence
of a delta function potential, we devise a technique of relating
the wave functions on either sides of these points (Appendix A).
\\

\noindent
Here we would like to mention that the SUSY method is a powerful
analytical method used for simple model potentials such as the
W-potentials. The applicability for calculating the Kramers'
escape rate in a W-potential for a {\it uniform} temperature
case has been demonstrated by Sch$\ddot{o}$nhammer \cite{schon}.
We found this method to be particularly useful while
investigating the influence of barrier subdivision on the escape
rate where we have demonstrated the existence of optimal number
of subdivisions that maximizes the escape rate \cite{mgn}.

\section{The model and its solution}

We consider a simple bistable potential in the form of a
symmetric W-potential which is piecewise linear having the same
magnitude in slope.  It is described by the barrier height $U_0$
and the distance $2L_0$ between the two potential minima located
at $x=\pm L_0$ on either side of the origin (Fig. 1a). The
localized hot zone of a certain width is taken to be positioned
somewhere between the two potential minima.  We assume a simple
temperature profile of the following form for the heat bath:
\begin{equation}
T(x)=T_0 + T_b\left[\Theta(x-d_b+\frac{w_b}{2}) -
\Theta(x-d_b-\frac{w_b}{2})\right].
\label{tmp}
\end{equation}
\noindent
Here, $\Theta(x)$ is the Heaviside function, $T_0$ is the
background (constant) temperature and $T_b$ is the excess
temperature of the hot zone.  The parameters $d_b$ and $w_b$
specify, respectively, the distance of the mid-point of the hot
zone from the barrier top and the width of the hot zone (Fig.
1b).\\

\noindent
The corresponding supersymmetric `partner' potential $V_-(x)$
with the hot zone is given by
\begin{equation}
V_-(x)=\frac{2u_0}{L_0}\left[ \delta(x+L_0) - \delta(x) +
\delta(x-L_0) \right] +
\left(\frac{u_{0,1}}{L_0}\right)^2.
\end{equation}
\noindent
This relation holds good at all points ${\it except}$ at the two
points of temperature discontinuity. Here, $u_0 =
\frac{U_0}{2T_0}$ and  $u_1 = \frac{u_0}{1 + s}$. It is clear that the 
potential $V_-(x)$ takes a constant value
$\left(\frac{u_0}{L_0}\right)^2$ {\it outside} the hot zone and
$\left(\frac{u_1}{L_0}\right)^2$ {\it within}. Two additional
repulsive and one attractive delta function potentials are
superposed at different locations as shown in Fig. 1c. Changing
the variable $x$ to $y = \frac{x}{L_0}$ gives a new
dimensionless Hamiltonian $h_-=L^2_0H_-$ such that
\begin{equation}
h_- \phi_-^0 (y) = e_b \phi_-^0 (y)
\end{equation}
\noindent
whose potential is
\begin{equation}
V_-(y) = 2 u_0 \left[ \delta (y+1) - \delta (y) + \delta (y-1) \right] +
(u_{0,1})^2.
\end{equation}
\noindent
Here, $e_b$ is a dimensionless quantity equal to $L^2_0 E_-$ or
equivalently $\frac{L_0^2\lambda_1}{\mu T_0}$. In the scaled
form, the position of the mid-point of the hot zone from the top
of the barrier and the width of the hot zone are defined by
$d=\frac{d_b}{L_0}$ and $w=\frac{w_b}{L_0}$, respectively (see
Fig. 1b). Henceforth, we use only the reduced variables.
Transfer matrix method has been used to find the
equation governing the eigenvalue $e_b$.  Below, we outline the
procedure leaving the details to the Appendix A.  \\

\noindent
The ground state wave function, $\phi_-^0$, is taken to have the
form $A_ne^{-k(y-y_n)} + B_ne^{k(y-y_n)}$ located with respect
to the positions $y=y_n$ of each of the delta function
potentials and at points of discontinuity in the temperature
profile.  The amplitudes $A_n$, $B_n$ and $A_n^{\prime}$,
$B_n^{\prime}$ on either sides of these special points are shown
in Fig. 1c. For regions outside the hot zone, the wave vector
$k$ is given by
\begin{equation}
k_0=\sqrt{u_0^2-e_b},
\end{equation}
\noindent
while for within the hot zone it is given by
\begin{equation}
k_1=\sqrt{u_1^2-\frac{e_b}{1+s}}.
\end{equation}
\noindent
The amplitudes $A_4^{\prime}$ and $B_4^{\prime}$ that appear
just on the left side of the positive delta function potential
located at $y=-1$ are related to $A_0$ and $B_0$ that appear on
the right side of the positive delta function potential located
at $y=1$ through a transfer matrix $\bf{M}$ such that
\begin{equation}
\left( \begin{array}{c}
A_4^{\prime}\\ B_4^{\prime}
\end{array}    \right) = \bf{M} \left( \begin{array}{c}
A_0\\ B_0
\end{array}  \right).
\end{equation}

\noindent
The matrix $\bf{M}$ is the product of successive transfer
matrices which are derived in the Appendix A. Since we are
looking for a bound state solution, $A_4^{\prime}=B_0=0$. This
implies that the 11-matrix element of ${\bf M}$ must be zero,
{\it i.e.},
\begin{equation}
M_{11}=0,
\label{root}
\end{equation}

\noindent
the solution of which gives the value of $e_b$ and, hence, the
equilibration rate.\\

\section{Results and discussion}

The effect of the hot zone on the equilibration rate for the given
potential is studied in terms of the three parameters
characterizing the hot zone: (i) the relative degree of hotness
of the hot zone with respect to the rest of the heat bath, 
$s=\frac{T_b}{T_0}$, (ii) the scaled width of the hot zone, 
$w=\frac{w_b}{L_0}$ and (iii) the scaled distance of the hot
zone from the top of the potential barrier, $d=\frac{d_b}{L_0}$.
Since the hot zone is in between the top of the barrier and the
right minimum, the parameters $d$ and $w$ can only take values
between 0 and 1.  On the other hand, the strength $s$ can take
positive values for hot zone and negative values (greater than
$-1$) for {\it cold} zone. In addition, for each of these cases
we have studied the influence of the barrier height
$u_0=\frac{U_0}{2T_0}$ on the equilibration rate.\\

\noindent 
The change in the equilibration rate due to the presence of the
hot zone is better appreciated in terms of a quantity which we
call {\it enhancement factor}, $f_b$, and defined by
\begin{equation}
f_b=\frac{e_b}{e_0}
\end{equation} 

\noindent
where $e_0$ stands for $e_b$ when there is no hot zone ($s=0$).
$f_b$ is nothing but the factor by which the equilibration rate
improves due to the presence of the hot zone.\\

\noindent
An expression for the enhancement factor in terms of all the
relevant parameters can be found from Eq.(\ref{root}). By noting
that, for a high barrier, $k_0$ and $k_1$ can be expanded about
$u_0$ and $u_1$, respectively, we find the enhancement factor to
be approximately given by
\begin{equation}
f_b=\frac{exp(s\sigma) \, cosh(s\sigma)} {1 + s \, exp(s\sigma)
\, sinh(\sigma) \, exp(-2u_0d)}.
\label{analy1}
\end{equation}
\noindent
where $\sigma\equiv\frac{wu_0}{1+s}$. Even though this equation gives
very accurate values comparable to those obtained by
numerically solving Eq.($\ref{root}$), the above equation can be
further approximated into a more transparent form when the hot
zone is located half way down the top ($d=0.5$) and is  given by 
\begin{equation}
f_b=\frac{1}{2} \left[1 + exp\left(2u_0w\frac{s}{1+s} \right)
\right].
\label{analy2}
\end{equation}

\noindent
Equation ($\ref{analy2}$) shows that
for a given barrier height $u_0$ and width $w$, $f_b$ increases
as the strength of the hot zone, $s$, increases and saturates
for large values.  In addition, the enhancement factor $f_b$ has
an exponential dependence on both the width $w$ of the hot zone
and the barrier height $u_0$. It may be noted that the above
expression has the correct limiting behaviour as a function of
both $s \rightarrow 0$ and $w \rightarrow 0$. Equation
($\ref{analy1}$) shows that the enhancement factor $f_b$
saturates as the blowtorch is moved away from the top of the
potential barrier.\\

\noindent
As an independent check of the analytical results 
(Eq.($\ref{analy1}$)), we have numerically solved Eq.($\ref{SE}$)
for a given temperature and potential parameters.  Before
numerically obtaining $f_b$, we have studied the evolution of
the probability distribution towards the steady state. This was
done by using an initial probability distribution sharply peaked
around the bottom of the right well.  A 3-dimensional plot of
the time evolution of the probability distribution towards the
steady state is shown in Fig. 2a. The numerically obtained
asymptotic steady state distribution is shown in Fig. 2b. This
clearly has higher population in the left well than the right
which in the absence of the hot zone would be symmetric about
the origin. Our numerical results show that this asymptotic 
distribution is practically identical
with the steady state distribution $P_{ss}(x)$ given by 
Eq.($\ref{Pss}$).\\ 

\noindent
One direct method of determining {\it the long time decay
rate} of the probability distribution is to allow the system to
evolve for long enough time towards the steady state. At late
stages when the probability distribution is observed not to show
any appreciable change, we can calculate the decay rate by
plotting the {\it total} probability on the right well as a
function of time on ($ ln \,\, t$ scale). However, it must be
mentioned here that for calculating the equilibration rate, a
more efficient way is to take the initial distribution to be a
distribution which is {\it marginally} different from the
stationary distribution (given by Eq.($\ref{Pss}$)). Even so,
high values of $u_0$ $(\sim 10)$ take long computer time for the
system to reach the near stationary state required for
calculating the equilibration rate. For this reason, for the
numerical works based on Eq.($\ref{SE}$) we have limited our
calculation to $u_0=4$.  Although this value of $u_0$ satisfies
the high barrier condition, it is on the low side.\\

\noindent
We now consider the influence of the hot zone on the
equilibration rate.  We shall refer to the results obtained from
the approximate analytic expression (Eq.($\ref{analy1}$)) as
analytical results, exact results obtained by numerically
solving the root of the equation (Eq.(\ref{root})) as
semi-analytical results and the results obtained by numerically
solving Eq.($\ref{SE}$) as numerical results.  Figure 3 shows
plots of the enhancement factor $f_b$ versus strength of the
blowtorch $s$ for various values of the barrier height $u_0$
placing the blowtorch (of width $w = 0.1$) midway from the top 
of the potential ($d=0.5$). (The set of curves $a$, $b$ and $c$
refer to three different values of the barrier height $u_0 =
4,10 $ and 15 respectively.) We have shown the
analytical results by a dashed line, the semi-analytical results
by a continuous line and the numerical results by filled
circles.  (For higher values of $u_0$, we have shown only the
analytical results and semi-analytical results, since numerical
results are time consuming.) As can be seen from Fig. 3, for
higher values of the barrier height $u_0$, analytical results
are not distinguishable from the semi-analytical results over
the entire range of $s$. This clearly reflects the fact that
large $u_0$ approximation made in obtaining Eq.($\ref{analy1}$)
holds good.  Consider the dependence of $f_b$ on the width of
the hot zone $w$. Figure 4 shows plots of $f_b$ as a function of
$w$ for $u_0 = 4, 10$ and $15$ labelled by $(a),(b)$ and $(c)$
respectively for $s=1$ and keeping the blowtorch half way down
the barrier top, {\it i.e.}, $d=0.5$. As before, for $u_0=4$, 
where we have numerical results, there is excellent agreement 
between the numerical results, and analytical and semi-analytical 
results over the entire range of $w$. For $u_0=10$
and $15$, semi-analytical results are indistinguishable from the
analytical results.  Lastly, we consider the dependence of the
enhancement factor, $f_b$, on the position of the hot zone from  
the barrier top, $d$. Figure 5 shows plots of $f_b$ versus $d$ for 
$u_0=4$, $10$ and $15$ denoted by $(a),(b)$ and $(c)$ respectively
(keeping the strength of the hot zone $s = 1$ and its width $w =
0.1$).  For relatively low barrier height, $u_0 = 4$, we see that the
analytical results agree well with the semi-analytical results and
the numerical results. Unlike for small barrier height for which $f_b$
saturates slowly, for large barrier heights ($(b)$ and $(c)$), $f_b$
saturates quickly and stays nearly constant beyond $d \sim 0.2$.
Here again, the analytical results agree very well with the
semi-analytical results. The above results reveal that the 
equilibration rate is higher when the hot zone is {\it away} from the 
barrier top.\\

\noindent
The above dependence of enhancement factor ($f_b$) for the
equilibration rate on the position of the blowtorch ($d$), shown
in Fig.5, is intriguing.  The enhancement factor is seen to
saturate quickly (to a value greater than unity) as the
blowtorch moves away from the barrier top ({\it i.e.}, as $d$
increases beyond $\frac{T_0}{U_0}$), while it  
decreases towards unity as the blowtorch 
approaches the top of the potential barrier. (Of course, for
sufficiently small $d$, the hot zone begins to bracket the
potential maximum because of finite width.) The above general
feature may be physically understood in terms of an argument due
to Landauer \cite{priv} based on the equilibration rate,
$\frac{1}{\tau} = \frac{1}{\tau_L} + \frac{1}{\tau_R}$, for the
occupation of the two wells, where $\tau_L$ and $\tau_R$ refer
to the time for crossing the barrier from the left and from the
right, respectively.  Thus, for a blowtorch on the right side of
the barrier, we expect not only $\tau_R$ to increase as we
approach the top of the barrier (as $d$ decreases), but also
$\tau_L$ should increase as the particles crossing the barrier
from the left well will be returned back by the blowtorch. This
might qualitatively explain the result of $d$ dependence of the
enhancement factor, $f_b$.  However, this argument does not
enable us to estimate the relative importance of the two
effects, namely, the variation of $\tau_L$ and $\tau_R$ with
$d$.  To clarify this point, we have carried out calculation of
$\tau$ for a highly asymmetric double-well potential in the
limit of a very deep right well. We considered two cases $-$ (i)
when the hot zone is placed to the right of the barrier top and
(ii) when it is to the left.  This enabled us to isolate to a
very good approximation the $d$ dependence of $\tau_L$ and
$\tau_R$ respectively from $\tau (d)$. {\em Our finding is that
the decrease in the enhancement factor, $f_b$, as $d$ decreases
is dominated by the increase of $\tau_R$, although ${\tau_L}$
also increases weakly}.  This is somewhat counterintuitive.

\section {Summary and conclusions}

In summary, we have been able to study the kinetic aspect of
Landauer's blowtorch theorem using the supersymmetric approach
and using a simple model $W$-potential.  The choice of the
$W$-potential is particularly well adapted to the SUSY method.
Our analysis shows that the rate of equilibration is
substantially improved due to the presence of the hot region.
The exact magnitude depends on its strength, its width, its
location and also on the barrier height. We have also obtained
an approximate analytical expression for the equilibration rate
from transfer matrices derived using the SUSY method. These
results agree well with the results obtained by numerical
solution of the associated Smoluchowski equation.\\

\noindent
We expect that the present analysis would be useful in
understanding some problems when local heating is viewed in a
more general context of local noise injection.  Viewed from this
angle, these results are clearly applicable even when
fluctuations are athermal.  One example where athermal
fluctuations play an important role is the depinning of
dislocation segments from obstacles resulting in movement of
dislocations \cite{mbthes}.  Another example where the present
analysis may be useful is the study of kinetics of phase
transformations.  In this case, the free energy takes the role
of the potential and the order parameter takes the role of the
reaction coordinate.  However, it is worth pointing out that in
general, subjecting a range of order parameter values to excess
heating is not easy since there is no correspondence between the
values taken by the order parameter and its location in space.
In this context, martensite transformation offers a promising
physical situation where we are actually dealing with the free
energy in terms of strain order parameter.  It would be
interesting to realize the applicability of the present analysis
to some experimental situations.\\

\section*{Acknowledgement}

\noindent
Authors would like to thank Prof. Rolf Landauer for very   
insightful comments. One of the authors 
(MB) would like to thank The International Program in Physical 
Sciences, Uppsala University, Uppsala, Sweden for financial 
support for this work. He would also like to thank Indian 
Institute of Science, particularly the Materials Research Centre 
Theory Group, for providing the necessary research facilities.\\

\newpage
\noindent
\section*{Appendix A}
In this appendix we outline the method of evaluating the matrix
${\bf M}$ that relates the amplitudes $A_4^{\prime}$,
$B_4^{\prime}$ of $\phi_-^0$ found on the left side of the delta
function potential at $y=-1$ to the amplitudes $A_0$, $B_0$
found on the right side of the delta function potential at
$y=1$.  (See Fig. 1c.) \\

\noindent
The region of the $W$ potential has four intervals of constant
potential $V_{-}(x)$ marked by I, II, III and IV as shown in
Fig. 1c.  In these intervals, only simple phase changes in the
wave function occur. In contrast, the wave function changes in a
significant way across regions of temperature discontinuity and
at points where delta functions are present.  First, consider
the phase changes which are simple to evaluate.  The pairs of
amplitudes $A_{n+1}$ and $B_{n+1}$ at the left end of these
regions(I to IV) are related to those at the right end,
$A_n^{\prime}$ and $B_n^{\prime}$, through the matrix equation 
\begin{equation}
\left( \begin{array}{c} 
A_{n+1} \\ B_{n+1}
\end{array}     \right)
= {\bf P}(a) \left( \begin{array}{c} A^{\prime}_n \\
B^{\prime}_n
\end{array}
\right)    
\end{equation}
\noindent
where
\begin{equation}
{\bf P}(a) = \left( \matrix{ e^{-ka} & 0\cr 0 & e^{ka}\cr}
\right).
\end{equation} 
\noindent
Here $a$ stands for the width of the concerned interval and $k$
is either $k_0$ or $k_1$ depending on whether the interval is
outside or within the hot zone.\\

\noindent
Now, consider relating the amplitudes across the points of
discontinuity in the temperature profile.  The probability
density $P(y,t)$ which is related to $\phi_-^0(y)$ through the
transformation Eq.($\ref{ansatz}$), is a continuous function of
the position {\it including} the two points where temperature is
discontinuous. Therefore, $\phi_-^0 (y)$ on either side of these
two points must be discontinuous in such a way that $P(y,t)$
remains continuous. In particular, at $y_1=d + \frac{w}{2}$,
using Eq.(3) and using the continuity condition
$P(y_1^+,t)=P(y_1^-,t)$, we get
\begin{equation}
\phi_-^0 (y_1^+)=\frac{\phi_-^0 (y_1^-)}{(1+s)}.\\
\label{phiy1a}
\end{equation}

\noindent
Here $y_1 ^\pm = y_1 \pm \epsilon$ with $\epsilon \rightarrow
0$.\\

\noindent
The second relation between the amplitudes comes from
integrating the SE equation, across $y_1$ and exploiting the
continuity of $P(y,t)$. This leads to the continuity of
$\frac{\partial}{\partial y} \left( T P \right) $.  At $y=y_1$,
this implies that
\begin{equation}
(\phi_-^{0} (y_1^+))^{\prime} = (\phi_-^{0} (y_1^-))^{\prime} ,
\label{phiy1b}
\end{equation}

\noindent
where the primes denote the spatial derivative.  Using Eqs.
($\ref{phiy1a}$) and ($\ref{phiy1b}$), the pairs of amplitudes
$A_1$, $B_1$ and $A_1^{\prime}$, $B_1^{\prime}$ found on the two
sides of the discontinuous temperature profile located at $y_1$
are related through the equation
\begin{equation}
\left( \begin{array}{c} 
A^{\prime}_1 \\ B^{\prime}_1
\end{array}     \right)
= {\bf M}_1 \left( \begin{array}{c} A_1 \\ B_1
\end{array}
\right)    
\end{equation}
\noindent
where
\begin{equation}
{\bf M}_1 = \frac{1}{2k_1} \left( \matrix{ (1+s)k_1 + k_0 &
(1+s)k_1 - k_0 \cr (1+s)k_1 - k_0 & (1+s)k_1 + k_0 \cr} \right).
\end{equation} 
\noindent
In the same way, the two sets of amplitudes $A_2$, $B_2$ and
$A_2^{\prime}$, $B_2^{\prime}$ found on the two sides of the
discontinuous temperature profile located at $y_2=d -
\frac{w}{2}$ are related through the matrix equation
\begin{equation}
\left( \begin{array}{c} 
A^{\prime}_2 \\ B^{\prime}_2
\end{array}     \right)
= {\bf M}_2 \left( \begin{array}{c} A_2 \\ B_2
\end{array}
\right)    
\end{equation}
\noindent
where
\begin{equation}
{\bf M}_2 = \frac{1}{2(1+s)k_0} \left( \matrix{ k_0 + (1+s)k_1 &
k_0 - (1+s)k_1 \cr k_0 - (1+s)k_1 & k_0 + (1+s)k_1 \cr}
\right).\\
\end{equation}

\noindent 
We are now left with relating the amplitudes on either sides of
the three delta function potentials located at $y=-1,0$ and $1$.
At all these three delta functions, $\phi_-^0$ is continuous
since temperature is continuous. Thus, the pairs of amplitudes
$A_0$, $B_0$ and $A_0^{\prime}$, $B_0^{\prime}$ on the right and
the left sides of the delta function located at $y=1$ are
related through
\begin{equation}
A_0^{\prime} + B_0^{\prime} = A_0 + B_0.
\end{equation}

\noindent
Integrating the Euclidean Schr$\ddot{o}$dinger equation across
this positive delta function with the width of the integration
tending to zero gives a second relation between these
amplitudes:
\begin{equation}
A_0^{\prime} - B_0^{\prime} = - \frac{2u_0}{k_0}\left( A_0 + B_0
\right) + A_0 - B_0 .
\end{equation}
\noindent
These two relations gives a matrix equation relating the two
sets of amplitudes:
\begin{equation}
\left( \begin{array}{c} 
A^{\prime}_0 \\ B^{\prime}_0
\end{array}     \right)
= {\bf M}_+ \left( \begin{array}{c} A_0 \\ B_0
\end{array}
\right).
\end{equation}
\noindent
The same transfer matrix ${\bf M}_+$ also relates the pairs of
amplitudes $A_4^{\prime}$, $B_4^{\prime}$ to $A_4$, $B_4$ since
features around $y=1$ are identical to those around $y=-1$.
When relating the pairs of amplitudes $A_3$, $B_3$ to
$A_3^{\prime}$, $B_3^{\prime}$ corresponding to the two sides of
the delta function located at $y=0$, we note that the only
difference is that the sign of the delta function is negative.
Going through the same procedure as above we get the matrix
equation:
\begin{equation}
\left( \begin{array}{c} 
A^{\prime}_3 \\ B^{\prime}_3
\end{array}     \right)
= {\bf M}_- \left( \begin{array}{c} A_3 \\ B_3
\end{array}
\right).    
\end{equation}
\noindent
The transfer matrices ${\bf M}_{\pm}$ are given by
\begin{equation}
{\bf M}_{\pm} = \frac{1} {k_0} \left( \matrix{ k_0 \mp u_0 &\mp
u_0\cr
\pm u_0        &k_0 \pm u_0
\cr} \right) 
\end{equation} 

\noindent
Using all the transfer matrices relating the successive sets of
amplitudes, the matrix {\bf M} is given by
\begin{equation}
{\bf M}= {\bf M}_+ {\bf P}(1) {\bf M}_- {\bf P}(d - \frac{w}{2})
{\bf M}_2 {\bf P}(w) {\bf M}_1 {\bf P}(1 - d - \frac{w}{2}) {\bf
M}_{+}.\\
\end{equation}

\noindent
\section*{Figure Captions}

\noindent
Figure 1: (a) The W - potential, (b) Temperature profile $T(x)$,
having the hot zone, (c) SUSY partner potential, $V_{-}(x)$.\\

\noindent
Figure 2: (a) Time evolution of the probability distribution
$P(x,t)$, towards the steady state for a given temperature.  The
parameter values are : $s=1.0$, $ u_0 = 4$, $w=0.1$ and $d=0.5$.
(b): Very late stage probability distribution obtained by
numerically integrating Eq.($\ref{SE}$) taken as the steady
state distribution.  Note the dip in the probability value at
the location of the hot zone shown by an arrow.\\

\noindent
Figure 3: Plot of $f_b$ versus $s$, for $d = 0.5, w = 0.1$ for
the values of (a) $u_0 = 4$, (b) $u_0 = 10$ and (c) $u_0 = 15$.
Dashed lines correspond to the results from Eq.($\ref{analy1}$), solid
lines to the semi-analytical method and filled circles to
numerical solution. \\

\noindent
Figure 4: Three plots of $f_b$ versus $w$, for $d=0.5, s=1$ for
the values of (a) $u_0 = 4$, (b) $u_0 = 10$ and (c) $u_0 = 15$.
Dashed lines correspond to the results from Eq.($\ref{analy1}$), solid
lines to the semi-analytical method and filled circles to
numerical solution. \\

\noindent
Figure 5: Plots of $f_b$ versus $d$ for $s=1.0$, $w=0.1$ for the
values of (a) $u_0 = 4$, (b) $u_0 = 10$ and (c) $u_0 = 15$.
Dashed lines correspond to the results from Eq.($\ref{analy1}$), solid
lines correspond to semi-analytical results and filled circles
to numerical solution.

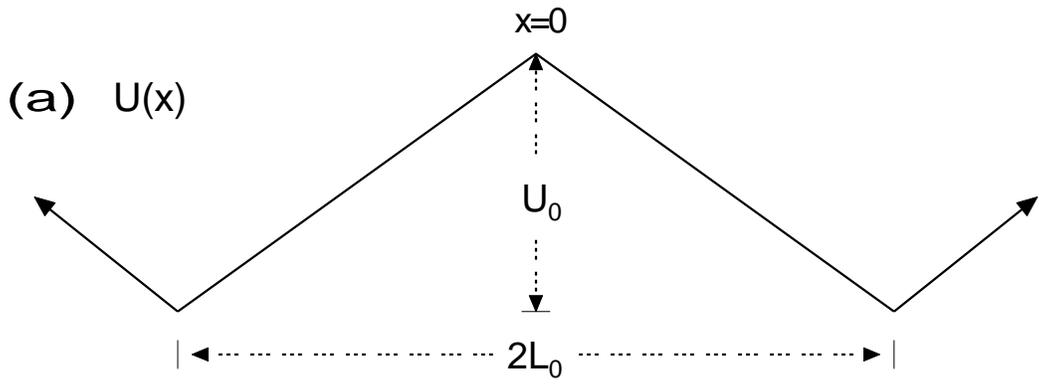
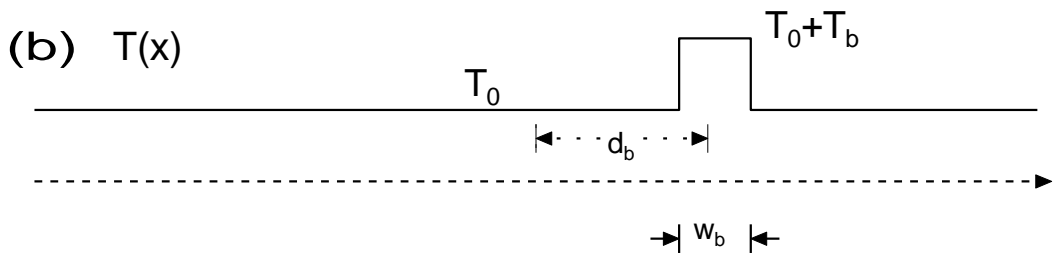
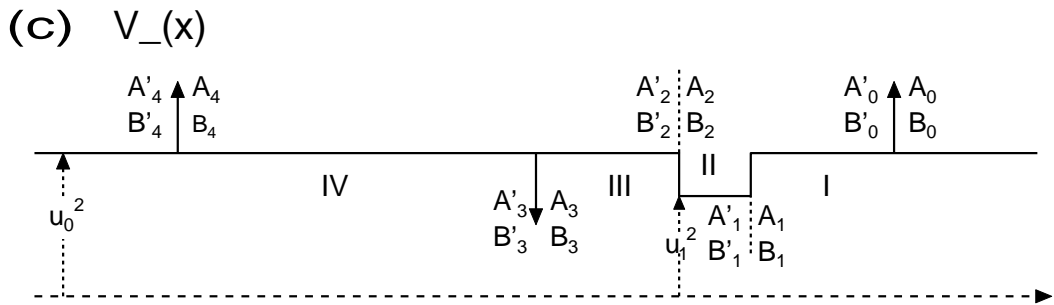
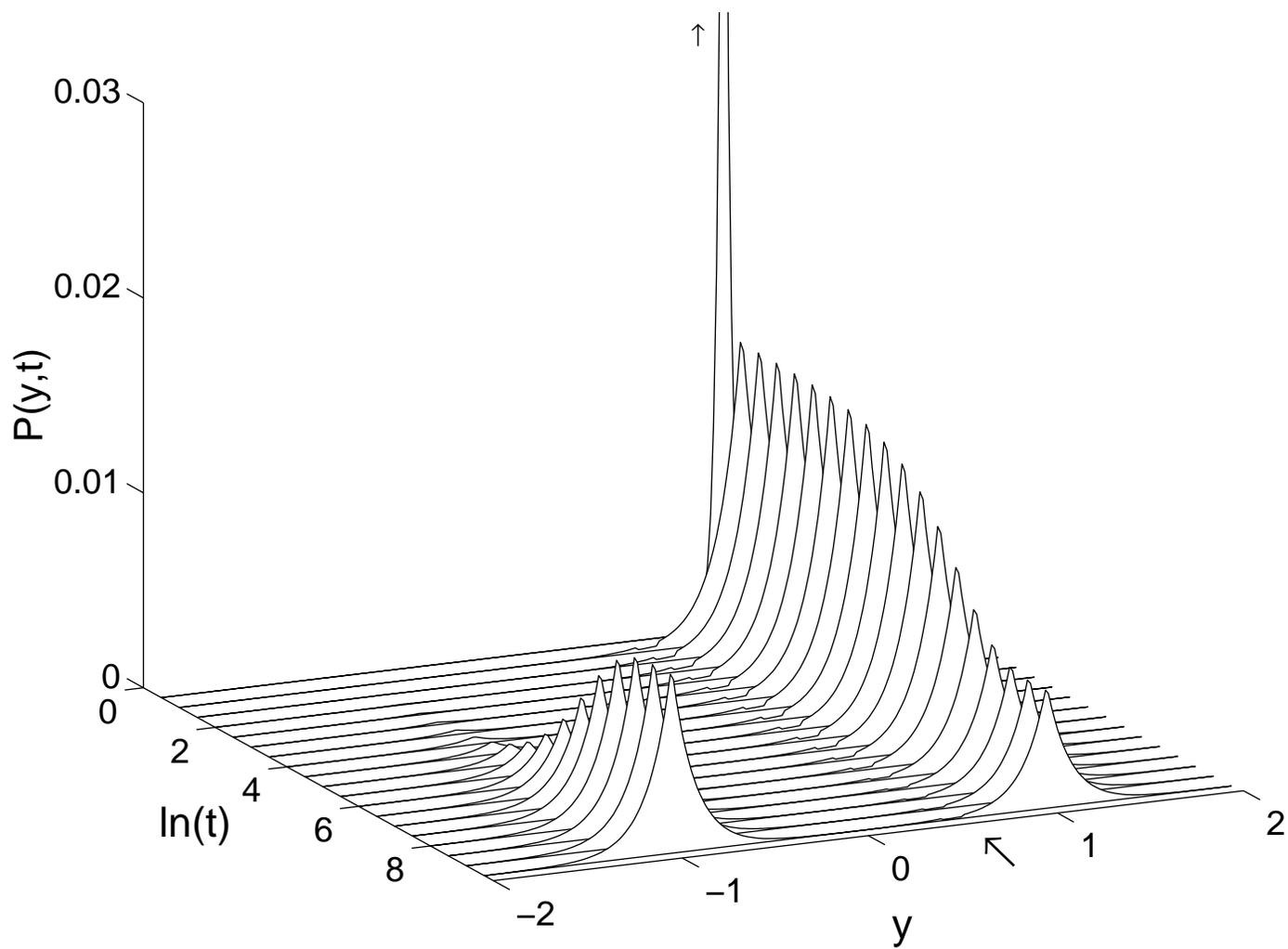
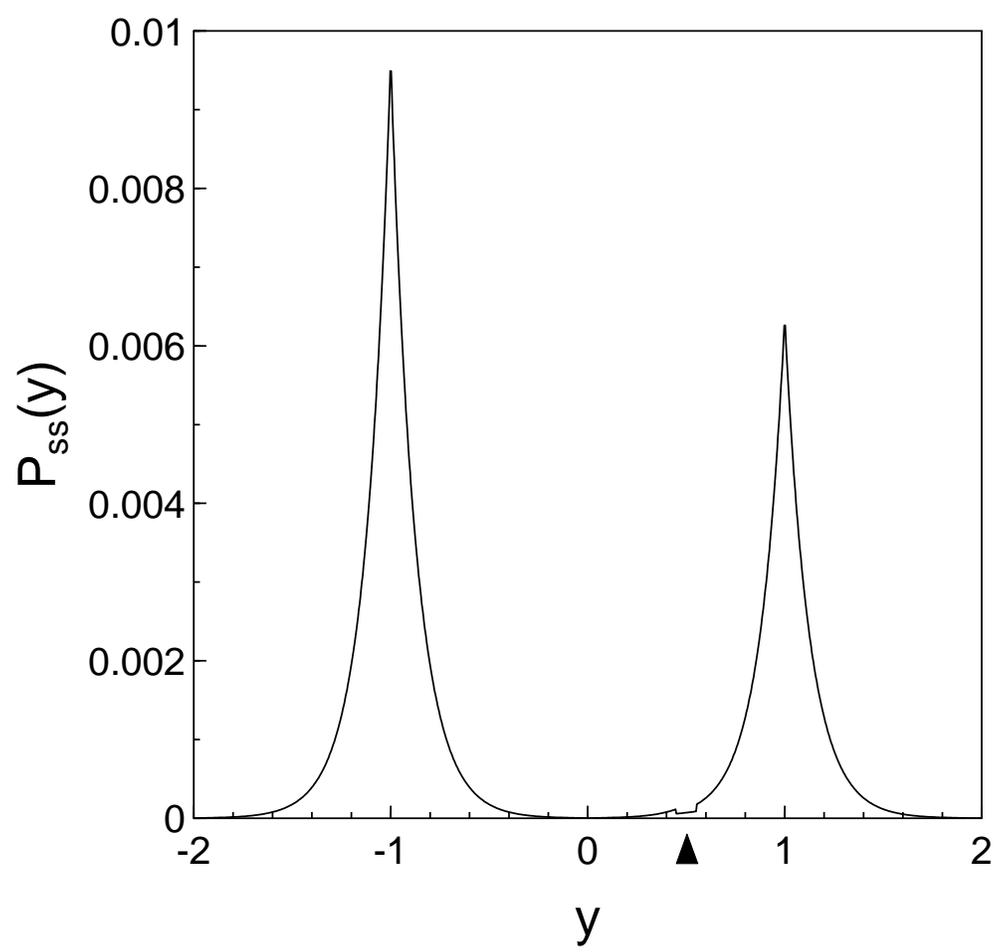
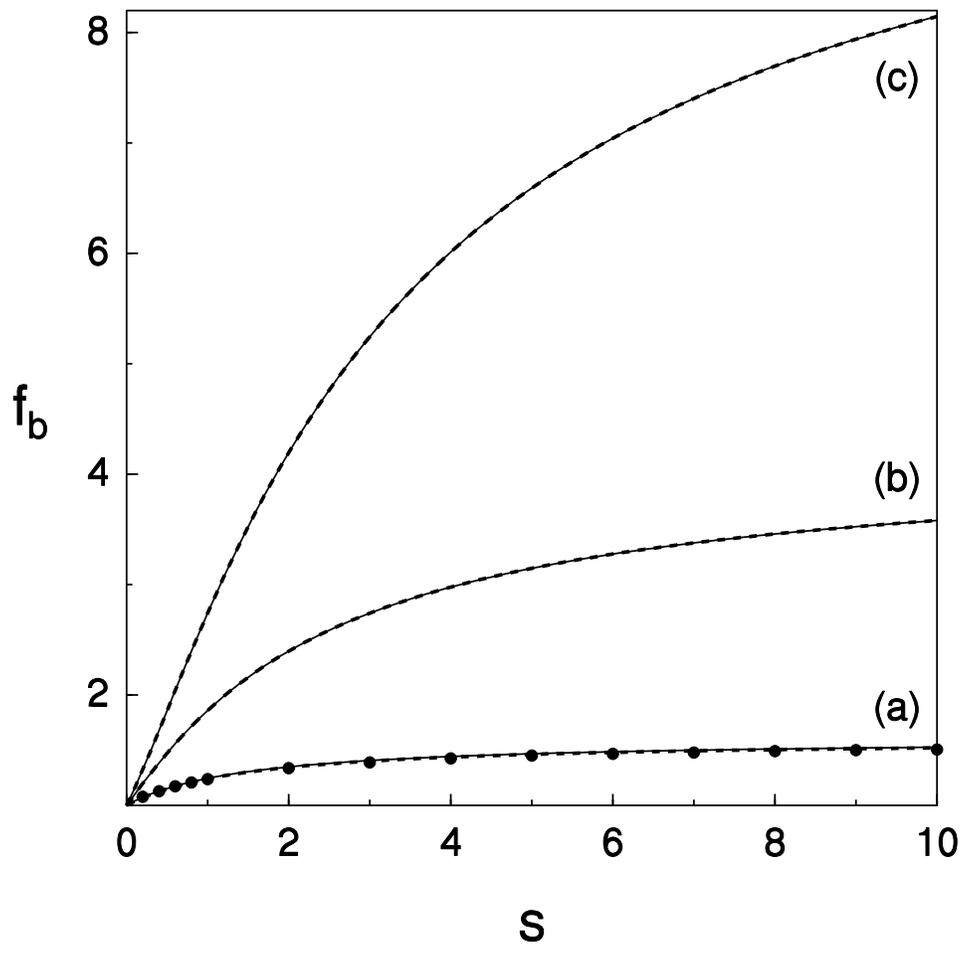
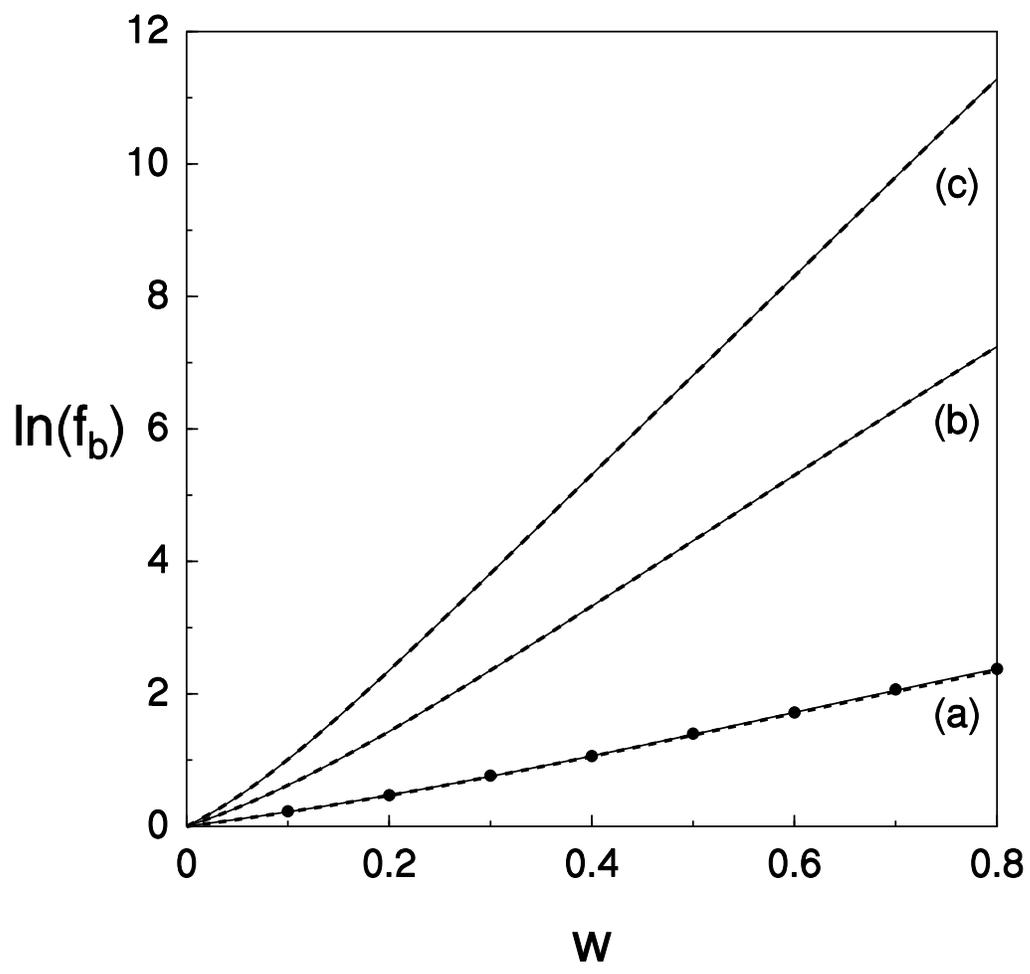
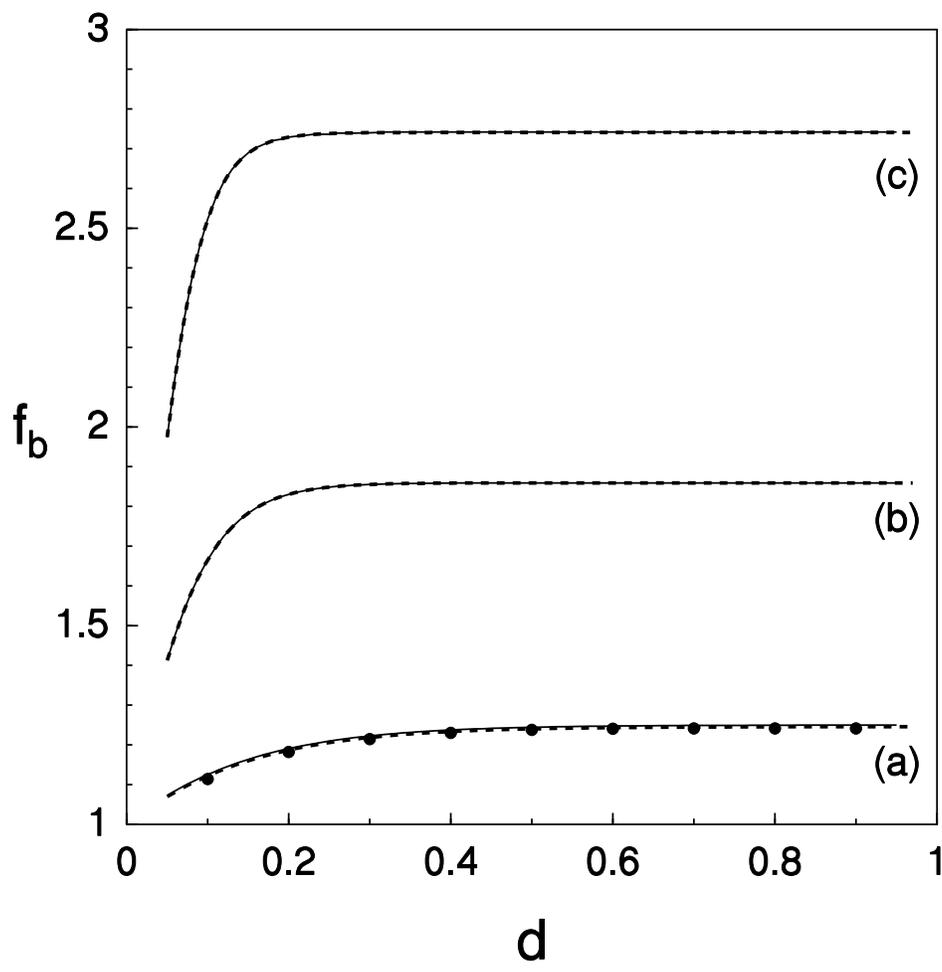
\end{document}